
\input phyzzx

\def\journal#1&#2(#3){\unskip, {#1}{\bf #2}(19#3),}
\def\andjournal#1&#2(#3){{#1}{\bf #2}(19#3),}
\def\andvol&#1(#2){{\bf #1}(19#2),}
\def\NP{Nucl. Phys. }

\def\PRL{Phys. Rev. Lett. }
\def\PL{Phys. Lett. }
\def\PTP{Prog. Theor. Phys. }

\def\Z{Z. Phys. }
\def\RMP{Rev. Mod. Phys. }
\def\j{\journal}

\REF\mizutori{
S.~Mizutori, Y.R.~Shimizu and K.~Matsuyanagi\j\PTP&83(90) 666;
\andvol&85(91) 559;
\andvol&86(91) 131;
in {\it Proceedings of the Workshop-Symposium on Future Directions in
Nuclear Physics with $4\pi$ Gamma Detection Systems of the New
Generation, Strasbourg, March 1991} (American Institute of Physics,
New York), p.287.}
\REF\nakatsukasa{
T.~Nakatsukasa, S.~Mizutori and K.~Matsuyanagi\j\PTP&87(92) 607.}
\REF\dudek{
J.~Dudek, T.R.~Werner and Z.~Szyma\'nski\j\PL&B248(90) 235.}
\REF\aberg{
S.~\AA berg\j\NP&A520(90) 35c.}
\REF\holler{
J.~H\"oller and \AA berg\j\Z&A336(90) 363.}
\REF\chasman{
R.R.~Chasman\j\PL&B266(91) 243.}
\REF\bonche{
P.~Bonche, S.J.~Krieger, M.S.~Weiss, J.~Dobaczewski, H.~Flocard and
P.-H.~Heenen\j\PRL&66(91) 876.}
\REF\li{
Xunjun~Li, J.~Dudek and P.~Romain\j\PL&B271(91) 281.}
\REF\nazarewicz{
W.~Nazarewicz and J.~Dobaczewski\j\PRL&68(92) 154.}
\REF\skalski{
J.~Skalski\j\PL&B274(92) 1.}
\REF\piepenbring{
P.~Piepenbring\j\NP&A541(92) 148.}
\REF\nazmitdinov{
R.~Nazmitdinov and S.~\AA berg\j\PL&B289(92) 238.}
\REF\skalskitwo{
J.~Skalski, P.-H.~Heenen, P.~Bonche, H.~Flocard and J.~Meyer, preprint
PNT/4192.}
\REF\cullen{
D.M.~Cullen et al.\j\PRL&65(90) 1547.}
\REF\sakamoto{
H.~Sakamoto and T.~Kishimoto\j\NP&A501(89) 205.}
\REF\kisslinger{
L.S.~Kisslinger and R.A.~Sorensen\j\RMP&35(63) 853}
\REF\brack{
M.~Brack, J.~Damgaard, A.S.~Jensen, H.C.~Pauli, V.M.~Strutinsky and
C.Y.~Wong\j\RMP&44(72) 320.}
\REF\nakatsukasatwo{
T.~Nakatsukasa, K.~Arita, K.~Matsuyanagi, S.~Mizutori and Y.R.~Shimizu,
to be published in {\it Proceedings of the International Conference on
Nuclear Structure at High Angular Momentum, Ottawa, May 1992}
(AECL-10613, Volume 2).}
\REF\semmes{
P.B.~Semmes, I.~Ragnarsson and S.~\AA berg\j\PRL&68(92) 460.}

\FIG\one{
Octupole strengths $|\bra{n}(r^3Y_{3K})^{''}\ket{0}|^2$ calculated for
the superdeformed states of $^{193}$Hg at $\omega_{\rm rot}=0$.
Note that positive- and negative-signature states are completely
degenerate at $\omega_{\rm rot}=0$ (for peaks with $K=1$, 2 and 3).
The deformation parameter $\delta_{\rm osc}=0.43$,
the neutron gap $\Delta_n=0.7$MeV, the proton gap $\Delta_p=0.7$MeV and the
doubly-stretched octupole interaction strengths
$\chi_{3K}=1.08\chi_{3K}^{\rm HO}$, $\chi_{3K}^{\rm HO}$ being the
selfconsistent values for the harmonic-oscillator potential, are used.
The numbers written beside the main peaks indicate the strengths for
the $E3$ operators measured in Weisskopf units.
}
\FIG\two{
The same as Fig. 1 but for $\omega_{\rm rot}=0.25$MeV$/\hbar$.
}
\FIG\three{
a) Quasiparticle energy diagram for neutrons with signature $\alpha=-1/2$
in $^{193}$Hg, plotted as a function of $\omega_{\rm rot}$.\nextline
b) The same as a) but the energy shifts ${\it\Delta}e'_{\rm vib}$
due to the coupling effects with the octupole vibrations are included.
Parameters used in the calculation are the same as in Figs. 1 and 2.
Notation like [512]5/2 indicate the main components of the wave
functions.
}
\FIG\four{
Amplitudes $C_0(\mu)$ and $C_1(\nu n)$ in the wave function defined by
Eq. (4), plotted as functions of $\omega_{\rm rot}$. The full lines are
used for the one-quasiparticle amplitudes, while the broken (dotted)
lines for the amplitudes involving the octupole vibrations with
positive (negative) signature. (a), (b), (c) and (d) respectively
show the results of calculation for bands 1, 2, $2'$ and 4.
The main component of band 2 is the [624]9/2($\alpha=1/2$) quasiparticle
state. The observed band 2 was suggested in Ref. \cullen ) that it could
actually be two bands with identical $\gamma$-ray energies consisting of
the [624]9/2($\alpha=1/2$) band and the [512]5/2($\alpha=1/2$) band.
The latter band, which is the signature partner of band 1, is denoted
here band 2$'$. The parameter used in the calculation are the same as
in Fig. 1.
}
\FIG\five{
Dependence of crossing frequency $\omega_{\rm cross}$ between bands 1 and
4, the aligned angular momentum of band 4 $i_{\rm band4}$, and the
interaction matrix element $V_{\rm int}$ between bands 1 and 4, on the
excitation energy $\hbar\omega_{K=2}^{(-)}$
of the lowest $K=2$ octupole vibration (with negative signature)
calculated at $\omega_{\rm rot}=0.45$MeV.
The pairing gaps used are the same as in Fig. 1. The excitation energy
$\hbar\omega_{K=2}^{(-)}=0.54$MeV corresponds to the force-strengths
$\chi_{3K}=1.08\chi_{3K}^{\rm HO}$.
}

\pubnum={KUNS 1165}
\date={October 1992}

\titlepage
\title{\bf
Effects of Octupole Vibrations on Quasiparticle Modes of Excitation
in Superdeformed $^{193}$Hg
}
\author{
Takashi~Nakatsukasa,
Shoujirou~Mizutori$^\ast$~and~Kenichi~Matsuyanagi
}

\address{
Department of Physics, Kyoto University, Kyoto 606-01
\break
$^\ast$Institute for Nuclear Study, University of Tokyo, Tanashi 188
}

\abstract{
A particle-vibration coupling calculation based on the RPA and the
cranked shell model has been carried out for superdeformed rotational
bands in $^{193}$Hg.
The result suggests that properties of single-particle
motions in superdeformed nuclei may be significantly affected by
coupling effects with low-frequency octupole vibrational modes,
especially by the lowest $K=2$ octupole mode.
}

\endpage
\hsize460pt
\hoffset0pt
\voffset-10pt
\vsize670pt
\sequentialequations

Since the shell structure of superdeformed nuclei is drastically
different from that of ordinary deformed nuclei, we expect that new kinds
of nuclear surface vibrational mode emerge above the superdeformed
yrast states. In fact, the RPA calculation in the uniformly rotating
frame, with the use of the single-particle states obtained by the cranked
Nilsson-Strutinsky-BCS procedure, has indicated that we can expect
highly collective, low-frequency octupole vibrational modes
(with $K=$0, 1, 2 and 3) about the superdeformed equilibrium
shape.$^{\mizutori ,\nakatsukasa )}$
Importance of the octupole correlations in superdeformed high-spin states
has been discussed also in Ref. \dudek )$\sim$\skalskitwo ).
The main reason why the
octupole is more favorable than the quadrupole is that each major shell
consists of about equal numbers of positive- and negative-parity
single-particle levels which are approximately degenerate in energy at
the superdeformed shape.

Existence of low-frequency octupole modes would imply that particle-hole
or quasiparticle modes of motion in superdeformed nuclei might be
significantly affected by the coupling effects with these vibrational
modes. In this paper, we report some results of theoretical calculation
which indicate the importance of such particle-vibration coupling effects
to understand the properties of Landau-Zener band-crossing phenomena
recently observed in $^{193}$Hg.$^{\cullen )}$

We solve the RPA equations for the Hamiltonian
$$
H=h'-{1\over 2}\sum_K \chi_{3K} Q^{''\dagger}_{3K}Q^{''}_{3K} \ ,
\eqn\hamiltonian
$$
where $h'$ is a cranked single-particle Hamiltonian of
the Nilsson-plus-BCS type,
$$
h'=h_{\rm Nilsson}-\Delta \sum_i (c_i^\dagger c_{\bar i}^\dagger
+c_{\bar i}c_i) -\lambda \hat N -\omega_{\rm rot}\hat J_x\ ,
\eqn\nilsson
$$
and $Q_{3K}^{''}=(r^3Y_{3K})^{''}$ are the doubly-stretched octupole
operators.$^{\sakamoto )}$
We  determine the equilibrium quadrupole deformation by means  of
the  Strutinsky method and use a large configuration  space
composed  of  9 major shells (for both protons  and  neutrons)  when
solving the coupled RPA dispersion equations.
The octupole-force strengths $\chi_{3K}$ can be determined by the
selfconsistency  condition between the density  distribution  and
the single-particle potential for the case of harmonic-oscillator
potential.$^{\sakamoto )}$
However, the problem how to generalize this method  to
a more general single-particle potential like Eq. \nilsson\ is  not
solved.     Therefore,     in     this     paper,     we      put
$\chi_{3K}=f\chi_{3K}^{\rm  HO}$, where $\chi_{3K}^{\rm HO}$  are
the theoretical values$^{\sakamoto )}$
for the harmonic-oscillator potential,  and
regard $f$ as a phenomenological parameter as well as the pairing
gap $\Delta$.

Figure 1 shows an example of the octupole strengths calculated at
$\omega_{\rm  rot}=0$  for the superdeformed $^{192}$Hg.  We  see
that  the  collectivity  is highest for  the  $K=2$
octupole  mode.  Figure 2 represents how  the  octupole  strength
distribution changes at a finite value of the rotational frequency
$\omega_{\rm  rot}$. In this figure, we can clearly  see  the
$K$-mixing  effects  due  to the Coriolis  force;  for  instance,
considerable  mixing among the $K=0$, 1 and 2 components is  seen
for the RPA eigenmode with excitation energy  $\hbar\omega\approx
1.0$MeV.

Starting from the microscopic Hamiltonian \hamiltonian\ and  using
the  standard  procedure,$^{\kisslinger )}$
we can derive the  following  effective
Hamiltonian   describing   systems  composed   of   quasiparticle
$a_\mu^\dagger$ and octupole vibrations $X_n^\dagger$,
$$
{\cal H}=\sum_\mu E_\mu a_\mu^\dagger a_\mu
+\sum_n \hbar\omega_n X_n^\dagger X_n
+\sum_n\sum_{\mu\nu}f_n(\mu\nu)(X_n^\dagger +\tilde X_n)
a_\mu^\dagger a_\nu\ .
\eqn\effective
$$
and    we    diagonalize    it   within    the    subspace
$\{a_\mu^\dagger\ket{0}$, $a_\nu^\dagger X_n^\dagger\ket{0} \}$.
The resulting state vectors can be written as
$$
\ket{\phi}=\sum_\mu C_0(\mu)a_\mu^\dagger\ket{0}
+\sum_n\sum_\nu C_1(\nu n)a_\nu^\dagger X_n^\dagger\ket{0}\ .
\eqn\wavefn
$$

Recently,  experimental data suggesting octupole correlations  in
superdeformed states have been reported by Cullen et al.$^{\cullen )}$
for $^{193}$Hg. Figure 3 shows a result of calculation for excitation
spectra  in the rotating frame of this nucleus. By comparing  the
conventional  quasiparticle energy diagram (Fig. 3-a))  with  the
result  of  diagonalization  of ${\cal H}$ (Fig. 3-b)),  we  can
clearly  identify  effects  of the  octupole  vibrations:  Energy
shifts  $\Delta e'_{\rm vib}$ of 50$\sim$300keV due to  the
coupling  effects are seen. In particular, we note that the
Landau-Zener crossing frequency $\omega_{\rm cross}$ between band 1
(whose main component is the [512]5/2 quasiparticle state) and band 4
(associated  with  the [761]3/2  quasiparticle)  is  considerably
delayed.    Namely,   we   obtain   $\omega_{\rm    cross}\approx
0.26$MeV$/\hbar$ in agreement with the experimental value$^{\cullen )}$
$\omega_{\rm rot}^{\rm exp}\approx 0.27$MeV,
whereas $\omega_{\rm  cross}\approx  0.17$MeV$/\hbar$  if  the
octupole-vibrational  effects are neglected.
The reason for this delay  is understood  by  examining the properties
of  the quasiparticle-vibration couplings in
$^{193}$Hg, which will be done below.

The amplitude $C_0(\mu)$, $C_1(\nu n)$ obtained by  diagonalizing
the  effective  Hamiltonian $\cal H$ are displayed  in  Fig. 4  as
functions  of the rotational frequency $\omega_{\rm  rot}$.  Note
that  the $K$-quantum numbers used in this figure to label  these
amplitudes are valid only in the limit
$\omega_{\rm  rot}\rightarrow 0$, because the $K$-mixing  effects
due to the Coriolis force are taken into account. It is seen that
the  main components of bands 1 and 4 exchange with each  other
at  $\omega_{\rm  rot}\approx  0.26$MeV$/\hbar$  indicating   the
Landau-Zener  crossing  phenomena between the  [512]5/2  and  the
[761]3/2  quasiparticle states. In this figure, we also see  that
the  mixing of the states composed of the [624]9/2  quasiparticle
and  the $K=2$ octupole vibration is significant in  band 1.  Note
that there are two such states;
$\ket{[624]9/2(\alpha=-1/2) \otimes \omega_{K=2}^{(+)} }$ and
$\ket{[624]9/2(\alpha=1/2) \otimes \omega_{K=2}^{(-)} }$
where $\alpha$ denotes the signature quantum number and
$\omega_{K=2}^{(+)}$ and $\omega_{K=2}^{(-)}$ the octupole vibrations
with positive and negative signatures, respectively, which reduce to
the $K=2$ octupole vibration shown in Fig. 1 in the limit of
$\omega_{\rm rot}=0$.

It is worth emphasizing that $K=2$ octupole matrix element between the
[512]5/2 and the [624]9/2 Nilsson states is especially large since it
satisfies one of the asymptotic selection rule
$({\it\Delta} N_{\rm sh}=1$, ${\it\Delta}n_3=1$, ${\it\Delta}\Lambda =2)$
for the transitions associated with the
$K=2$ octupole operator.
($N_{\rm sh}$ denotes the shell quantum number,
defined by $N_{\rm sh}=2(n_1+n_2)+n_3$.$^{\nakatsukasa )}$)
Thus, these two Nilsson states strongly couple
with each other due to the $K=2$ octupole correlation. This property is
seen also in the single-neutron energy diagram plotted as a function of
the $K=2$ octupole deformation parameter $\beta_{32}$ in the paper by
Skalski.$^{\skalski )}$
As a result of this property, we obtain a significant energy
shift ${\it\Delta} e'_{\rm vib}$ for band 1. On the other hand, the
octupole vibratinal effect is rather weak for band 4. Thus, as shown in
Fig. 3, the relative excitation energy between bands 1 and 4 increases,
so that their crossing frequency also increases.
Furthermore, we can expect that this octupole correlation between
bands 1 and 2 may contribute to the
strong ($E1$) transitions from band 1 ([512]5/2) to band 2 ([624]9/2),
which have been also reported in the experiment.$^{\cullen )}$

Next, let us discuss on the alignment $i$ of band 4 and on the
interaction matrix element $V_{\rm int}$ between bands 1 and 4, for which
experimental  data are available;
$i_{\rm band4}^{\rm exp}\approx 1.3\hbar$ and
$V_{\rm int}^{\rm exp}\approx 26$keV.$^{\cullen )}$
We evaluate the alignment by
$i=-\partial E' /\partial\omega_{\rm rot}$ using the eigenvalue $E'$
of the effective Hamiltonian \effective\ and choosing the region of
$\omega_{\rm rot}$ where $E'$ linearly depends on $\omega_{\rm rot}$.
The interaction $V_{\rm int}$ is evaluated, as usual,
from the half of the
shortest distance between the energy levels for bands 1 and 4 in the
energy diagram like Fig. 3-b). The calculated value of the alignment for
the [761]3/2 quasiparticle state (the main component of band 4) is
$i^{\rm cal}\approx 1.8\hbar$. This value is reduced to
$i^{\rm cal}\approx 1.2\hbar$ in good agreement with experiment, when
the octupole-vibrational effects are taken into account.
On the other hand, the interaction matrix element between the [761]3/2
quasiparticle state and the [512]5/2 quasiparticle (the main component
of band 1) is almost zero and increases to about 5keV due to  the
octupole-vibrational effects. This calculated value of $V_{\rm int}$
is, however, too small in comparison with the experimental data.

Since we treat the doubly-stretched octupole force-strengths $\chi_{3K}$
as phenomenological parameters in this paper, it is necessary to examine
the dependence on the force-strengths $\chi_{3K}$, of the theoretical
values for the crossing frequency $\omega_{\rm cross}$, the alignment
$i_{\rm band4}$ and the interaction matrix element $V_{\rm int}$.
This is done in Fig. 5. In this figure, the calculated values of
$\omega_{\rm cross}$, $i_{\rm band4}$ and $V_{\rm int}$ are plotted as
functions of the excitation energy $\hbar\omega_{K=2}^{(-)}$
of the lowest $K=2$ octupole vibration calculated at
$\omega_{\rm rot}=0.45$MeV$/\hbar$, instead of plotting directly as
functions of $\chi_{3K}$. We note that $\hbar\omega_{K=2}^{(-)}$ is a
function of $\chi_{3K}$ and the force-strengths
$\chi_{3K}=1.08\chi_{3K}^{\rm HO}$ adopted in the calculations of
Figs. 1$\sim$4 correspond to the abscissa at $\hbar\omega_{K=2}^{(-)}
\approx 0.5$MeV in Fig. 5. It is seen from this figure that
$\omega_{\rm cross}$ increases while $i_{\rm band4}$ decreases when
$\hbar\omega_{K=2}^{(-)}$decreases (\ie , when the octupole-vibrational
effects become stronger), and we find that the experimental data for
$\omega_{\rm cross}$ and $i_{\rm band4}$ are simultaneously reproduced
at $\hbar\omega_{K=2}^{(-)}\approx 0.5$MeV. On the other hand, the
calculated interaction matrix element $V_{\rm int}$ is too small within
a reasonable range of $\hbar\omega_{K=2}^{(-)}$.

The main reason why the calculated value of $V_{\rm int}$ is small
may be understood as follows: Generally speaking, we can expect that
the  band-band interactions increase due to  the  octupole-vibrational
effects, because interactions between different quasiparticle states
through intermediate configurations composed of one-quasiparticle and
octupole vibrations become possible. However, in the specific case of
band 4, as seen in Fig. 4, the octupole vibrational effects are rather
weak because the octupole matrix  element between the [761]3/2 Nilsson
state and the neighboring Nilsson states are small. On the  other
hand,
the octupole-vibrational effects are indeed strong in band 1 so that the
considerable mixing of the states
$\ket{[624]9/2(\alpha=\mp 1/2)\otimes\omega_{K=2}^{(\pm )}}$ occurs.
This mixing does not, however, lead to a large interaction $V_{\rm int}$
between bands 1 and 4, because the octupole matrix element between
the  [624]9/2 and the [761]3/2 quasiparticle states remains small
although the Coriolis $K$-mixing effects are taken into account
in our calculation  at finite rotational frequency.

The interaction $V_{\rm int}$ under consideration depends also on the
pairing-gap parameter $\Delta$ as well as the force-strengths
$\chi_{3K}$. We have adopted $\Delta=0.7$MeV in Fig. 5 (cf. we obtain
$\Delta_p=0.72$MeV and $\Delta_n=0.77$MeV when
the pairing gap is evaluated  at $\omega_{\rm rot}=0$ by means of the
conventional procedure of the Strutinsky method$^{\brack )}$
with the pairing-force
strengths $G$ that gives the standard value of the smoothed pairing-gap
parameter $\tilde\Delta=12.0A^{-1/2}$MeV).
The result of calculation using
$\Delta=0.9$MeV was reported in Ref. \nakatsukasatwo ).
In this case, we obtain
$V_{\rm int}\approx 10$keV keeping the agreement of $\omega_{\rm cross}$
and $i_{\rm band4}$ with experiment. This value of $V_{\rm int}$ is
still too small compared with the experimental value
$V_{\rm int}\approx 26$keV. Thus, we conclude that the large value of
$V_{\rm int}$ cannot be reproduced within the present framework of
calculation by merely changing the pairing-gap parameter $\Delta$
within a reasonable range.

In summary, we have investigated the coupling effects between the
quasiparticle and the octupole-vibrational modes of excitation in the
superdeformed $^{193}$Hg, by means of the particle-vibration coupling
theory based on  the cranking model. We have found that the inclusion
of the octupole vibrational effects is important to reproduce the
experimental data for the crossing frequency between bands 1 and 4,
and for the aligned angular momentum of band 4. On the other hand, the
calculated interaction matrix element between bands 1 and 4 is too
small in comparison with the experimental data. To understand the
spectrum of the superdeformed $^{193}$Hg, there are several problems
remaining for the future; e.g., improvement of the treatment of the
pairing correlations, inclusion of the quadrupole-pairing, evaluation
of the $E1$ transition probabilities,
possibilities of other interpretation of the experimental
data,$^{\semmes )}$ etc.

The computer calculation for this work has been financially supported
in part by Research Center for Nuclear Physics, Osaka University and
by Institute for Nuclear Study, University of Tokyo.
\endpage
\refout
\figout

\bye